# Medical Isotope Production with the IsoDAR Cyclotron

Jose R. Alonso[1,*], Roger Barlow[2], Janet M. Conrad[1], and Loyd Hoyt Waites[1]

[1]Massachusetts Institute of Technology, Cambridge, MA 02139, USA

[2]The University of Huddersfield, Queensgate, Huddersfield, HD1 3DH, UK

* Email: jralonso@lbl.gov

**Standfirst:** Jose R. Alonso and colleagues describe technical advances that will allow the proposed IsoDAR cyclotron — developed for neutrino physics research — to produce medical isotopes more efficiently than existing cyclotrons can.

**Introduction**

A wide variety of clinical isotopes can be produced by cyclotrons[1], by bombarding other atoms with accelerated particles. Example isotopes include $^{19}$F used in fluorodeoxyglucose for positron emission tomography (PET), $^{99m}$Tc (which currently is primarily reactor-produced) used for cardiac stress tests, and $^{131}$I used to treat hyperthyroidism. Each application has its own specific biology and chemistry; for each isotope, the beam energy, beam particle, and target are also unique. Although the production of many radioisotopes is established, there are avenues for improvement. For example, it could be more efficient and replace the dwindling number of research reactors with cyclotron-based facilities, or it could be enhanced for the production of isotopes whose small cross sections make their use presently uneconomical. Enabling these improvements requires an increase in the current delivered by the cyclotron, which for existing machines is typically no more than 1 mA.

The proposed IsoDAR cyclotron, to be built at the Kamioka Observatory in Japan, is a compact cyclotron designed primarily for sterile neutrino searches. IsoDAR will deliver 10 mA of 60 MeV protons[2], an order of magnitude higher than existing cyclotrons. It is designed[3] to place a powerful neutrino source in close proximity to a large liquid scintillator detector (such as KamLAND) as a definitive test for the existence of sterile neutrinos. Its 600 kW of protons strike a Be target to produce neutrons that flood a $^{7}$Li sleeve generating $^{8}$Li, whose decay produces electron antineutrinos with a flux equivalent to that of a 50 kiloCurie (2 petaBecquerel) beta-decay source. The cross section for neutrino interactions is so small that to obtain a statistically meaningful number of observed interactions a very large number of neutrinos must be produced, which in turn requires accelerator technology well beyond today's limits. This dramatic improvement of performance, driven by the requirements of high-energy physics, can be put to use in opening new possibilities for the production of medical isotopes.

**Cyclotron technology**

Cyclotrons use radio frequency (RF) electric fields to accelerate charge particles, which are held on spiral trajectories by static magnetic fields. The first cyclotron was built by Ernest Lawrence in 1930 as a tool for nuclear physics research. As the field developed, nuclear physicists — and later, particle physicists — required ever-increasing beam energies for basic physics studies. Because the cyclotron can only produce lower-energy particles, restricting it to MeV rather than GeV or TeV energies, it was replaced by the synchrotron as the machine for science at the frontiers of high energy. However, the MeV energies attainable by the cyclotron are ideal for producing radioisotopes.

Early cyclotrons accelerated protons, the simplest ion. However, the current of a proton beam is limited by the process of extracting the beam from the magnet. This extraction is done using a septum: a thin sheet placed between the last and second-last orbital turns, which allows the establishment of an electric or magnetic field that deflects particles in the last turn out of the magnet and towards the desired target. However, if the pitch of the spiral trajectory is not large enough, some protons will strike the septum plate, causing activation and damage. Today's cyclotrons use $H^-$ ions, which have the same dynamics as protons (apart from a minus sign), but which can be extracted using a very thin stripping foil. This foil removes the two electrons from the ion leaving a proton, which will bend the opposite way in the field. Although using a stripper foil rather than a septum increases the possible machine current, there are still two important limitations: stripper-foil lifetime and space-charge-induced beam blowup. Heat deposited in the foil from the passage of ions limits $H^-$ currents to a maximum of about 1 mA. The electrons released by the break-up of the $H^-$ ion on the foil are bent with very small radii and deposit all their energy through repeated passes through the foil. Over 90% of the foil heating comes from these electrons. Currents are also limited by space charge — the repulsive force between like charges which causes a bunch of particles to expand, resulting in beam loss on the vacuum-tank surfaces.

The IsoDAR cyclotron will accelerate $H_2^+$ instead of $H^-$, a choice which has several advantages[4]. One advantage is that although septum extraction would be the primary means of taking the beam out of the cyclotron, beam extraction can be aided by stripper foils, even at the higher current levels. The $H_2^+$ ion passing through a foil generates only one free electron, which is bent away from the magnet centre, and can be intercepted by a catcher before re-entering the foil. (Such a catcher cannot be used with $H^-$ because it would need to be on the inner radius, and would interfere with inner turns.) The now-free proton has a much smaller bending radius, and can be made to spiral around, using the field differences between hill and valley regions of the cyclotron magnet, to exit cleanly from the cyclotron. A very narrow stripper foil can be used to shadow the septum plate, intercepting ions that would strike it. The protons emerging from this stripper foil are bent inwards, and avoid the septum. In addition, stripper foils could be used to extract larger portions of the beam directly into production targets, becoming the primary extraction technique. A $H_2^+$ beam also reduces the effect of space charge in two ways: the $H_2^+$ ion has two protons for every charge (effectively doubling the proton current), and doubling the ion mass reduces the kinematic growth of the bunch from the electrostatic repulsion forces.

The choice of $H_2^+$ also means that the cyclotron has the flexibility to accelerate other ions with the same charge-to-mass ratio, such as deuterons, alpha particles or $C^{6+}$. This opens possibilities for generating isotopes not accessible with proton beams.

Beyond accelerating $H_2^+$, another major innovation of the IsoDAR cyclotron is the use of a radio frequency quadrupole (RFQ) to pre-bunch the beam into the cyclotron, which substantially improves efficiency over other cyclotrons. For example, the injection line in a compact cyclotron brings the continuous beam from the ion source along an axial channel into the centre of the cyclotron, where it is bent into the plane of the magnet. The RF accelerating system (known as Dee's for cyclotrons) captures ions that are in the correct phase, which is typically only about 10% of the incident beam. 'Traditional' RF bunchers improve this efficiency to about 20%. This inefficiency requires more powerful ion sources, and the large beam losses in the central region cause heat and erosion damage to the cyclotron. A properly designed RFQ can increase the capture efficiency to 80% or more. However, the RFQ must operate at the cyclotron frequency, of the order of 30 MHz, which is exceptionally low for traditional RFQs. Recent designs, using 4-rod geometries, however, can meet this requirement. The 'front end' of the IsoDAR cyclotron, including an ion source optimized for $H_2^+$ and the RFQ is currently under construction[5].

The table shows a comparison between IsoDAR and two widely-used commercial cyclotrons. IsoDAR is somewhat larger and heavier, because of the higher momentum required to achieve an equivalent energy per proton for the $H_2^+$ ion, but this is a modest price to pay for the 10-fold increase in available current.

**Isotope production**

We illustrate the possibilities opened up for isotope production with two examples: $^{68}$Ge and $^{225}$Ac. The $^{68}$Ge / $^{68}$Ga generator is receiving a lot of attention in the isotope community now, as an effective means of generating isotopes for PET. Efficient isotope generation for PET may fill the gap that would be caused by a reduction in access to radioactive tracers such as $^{99m}$Tc, which may be subject to unreliable access to $^{99}$Mo / $^{99m}$Tc generators in coming years. The IsoDAR cyclotron could increase the production of $^{68}$Ge by as much as a factor of 10 over the IBA C-30 cyclotron. The factor of 5 increase in current is supplemented by the higher energy of the beam on target, allowing effective use of both of the naturally-occurring isotopes of the Ga target material. Passing the 60 MeV beam through a thick target of natural Ga (40% $^{71}$Ga and 60% $^{69}$Ga ) would first access the $^{71}$Ga(p,4n) $^{68}$Ge reaction, then, deeper in the target, the $^{69}$Ga(p,2n) $^{68}$Ge reaction. Both reactions proceed via the compound nucleus channel, with about equal cross sections (~150 millibarn). The (p,4n) reaction peaks at about 50 MeV, the (p,2n) at 25 MeV, both excitation functions being about 15 MeV wide.

The second example is $^{225}$Ac, a highly promising alpha emitter that has shown high effectiveness in therapy applications[6]. Early production techniques employed extremely complex target preparation and chemical separation processes, leading to very high costs and very limited access. In the US, Los Alamos National Laboratory and Brookhaven National Laboratory have developed a technique involving the irradiation of a natural Th target[7], with quite high cross sections at 200 MeV but still reasonably good at 60 MeV. They estimate being able to use the facilities at BLIP and LANSCE to provide a factor of 60 more $^{225}$Ac per year than is currently available. The IsoDAR cyclotron could increase this amount by another factor of 100.

In summary, we believe that the performance breakthroughs developed for the IsoDAR cyclotron can have significant benefits for isotope production. These developments are presented in greater detail in a separate publication[2]. One possible challenge for IsoDAR is that the beam power available far exceeds present-day target handling capabilities. However, techniques exist for splitting the beam between many targets; furthermore, the availability of the higher power will surely stimulate the development of targets capable of handling higher powers. Although IsoDAR is larger and hence more expensive than existing cyclotrons, the huge increase in yield of difficult-to-produce isotopes should provide a rationale for inclusion of one or more of these cyclotrons in major isotope-production centres. The high yield also requires sophisticated target handling and processing equipment not likely found in a hospital, again suggesting that large commercial centres should be the site for these sophisticated accelerators. IsoDAR joins superconducting magnets, the World Wide Web and its high-capacity data-handling systems, and many other technology innovations developed to solve problems in particle physics, that are now applied in many diverse fields.

Table 1: Comparison of IsoDAR with IBA commercial isotope cyclotrons

| Parameter | IsoDAR | IBA C-30 | IBA C-70 |
| --- | --- | --- | --- |
| Ion species accelerated | $H_2^+$ | $H^-$ | $H^-$ |
| Maximum energy (MeV/Da) | 60 | 30 | 70 |
| Proton beam current (mA) | 10 | 1.2 | 0.75 |
| Available beam power (kW) | 600 | 36 | 52 |
| Pole radius (m) | 1.99 | 0.91 | 1.24 |
| Outer diameter (m) | 6.2 | 3 | 4 |
| Iron weight (ton) | 450 | 50 | 140 |
| Electric power required (MW) | 2.7 | 0.15 | 0.5 |

The final published version of this Comment is available at https://www.nature.com/articles/s42254-019-0095-6
This work was supported by the US National Science Foundation, Grant NSF-PHY-1505858.